\documentclass[3p,times,procedia]{elsarticle}
\usepackage{nupha_ecrc}

\volume{00}

\firstpage{1}

\journalname{Nuclear Physics A}

\runauth{Sedigheh Jowzaee for the STAR Collaboration}

\jid{nupha}

\jnltitlelogo{Nuclear Physics A}

\usepackage{amssymb}

\usepackage{lineno}

\usepackage[figuresright]{rotating}

\begin{document}

\begin{frontmatter}



\dochead{}

\title{Rapidity correlations in the RHIC Beam Energy Scan Data}


\author{Sedigheh Jowzaee for the STAR Collaboration\fnref{1}}

\address{Wayne State University, 666 W. Hancock, Detroit, MI 48021}
\fntext[1]{A list of members of the STAR Collaboration and acknowledgements can be found at the end of this issue.}

\begin{abstract}
A pair-normalized two-particle covariance versus the rapidity
of the two particles, called R$_2$, was originally studied in ISR 
and FNAL data in the 1970's. This variable has recently seen
renewed interest for the study of the dynamics of heavy-ion
collisions in the longitudinal direction. These rapidity
correlations can be decomposed into a basis set of Legendre
polynomials with prefactors $\langle a_{mn}\rangle$,
which can be considered the rapidity analog of the
decomposition of azimuthal anisotropies into a set of
cosine functions with prefactors v$_{\rm n}$. The
$\langle a_{mn}\rangle$ values have been suggested
to be sensitive to the number of particle emitting sources, baryon stopping,
viscosities, and critical behavior. The rapidity correlations have
been measured by the STAR collaboration as a function of the
beam energy for 0-5\% central Au$+$Au collisions 
with beam energies ranging from 7.7 to 200~$\mathrm{GeV}$.
The experimental results and comparisons to the
UrQMD model are presented.
\end{abstract}


\end{frontmatter}


\section{Introduction}
Initial state density fluctuations and the pressure gradient of the strongly interacting medium in transverse plane result in azimuthal anisotropies of the emitted particles in the final state. These can be characterized by a Fourier expansion with magnitudes given by prefactors v$_{\rm n}$ \cite{1,2}. 
The longitudinal structure of the initial state, particularly at RHIC energies, remains largely unexplored. The density fluctuations can be studied in terms of correlations of particles in the longitudinal direction, {\it i.e.} in (pseudo)rapidity \cite{3,4}. 

Long-range rapidity correlations, and other ``short range" mechanisms such as resonance decays, jet fragmentation, and quantum statistical effects, appear in specific ways in the correlation functions \cite{4}. The shape of the longitudinal correlations can be characterized by Chebyshev \cite{3} or Legendre polynomials \cite{4} with magnitudes given by prefactors $a_{nm}$. 
Recently, the ATLAS collaboration has studied \cite{5} the longitudinal fluctuations in terms of $a_{nm}$ and provided information on the early time dynamics in different collision systems. A relativistic hydrodynamical calculation \cite{6} has also been used to show that the rapidity correlations are sensitive to the number of particle-producing sources and the transport properties of the medium. The importance of doing this analysis at RHIC was also emphasized in Ref\cite{6}.

In this work, rapidity correlations were studied with STAR for Au$+$Au collisions at eight different beam energies from 7.7 to 200~$\mathrm{GeV}$ for the BES-I program at RHIC. The calculated Legendre coefficients $\langle a_{nm}\rangle$ from experimental data are also compared with the results from the UrQMD model \cite{6b}.


 \section{Analysis Method}
Two-particle rapidity correlation function is defined as \cite{7}:
\begin{equation}
       R_{2}(y_{1},y_{2}) =\frac{\langle\rho_{2}(y_{1},y_{2})\rangle}{\langle\rho_{1}(y_{1})\rangle\langle\rho_{1}(y_{2})\rangle}-1 ;
\label{formula1}
\end{equation}
where $\langle\rho_{2}(y_{1},y_{2})\rangle$ and $\langle\rho_{1}(y_{1,2})\rangle$ are two particle, and single particle, multiplicity density distributions in rapidity, respectively; averaged over events within a narrow centrality class. 
In order to study the purely dynamical fluctuations, a normalization is used to decouple the residual centrality dependence from the rapidity correlation function, $C(y_{1},y_{2})$=$R_{2}(y_{1},y_{2})$+1 \cite{4,5}. This normalized correlation function is given by:
\begin{equation}
       C_{N}(y_{1},y_{2}) =\frac{R_{2}(y_{1},y_{2})+1}{C_{p}(y_{1})C_{p}(y_{2})} ;
\label{formula2}
\end{equation}
where $C_{p}$ is the projection of the correlation function along the rapidity axis of each particle.
The correlation function $C_{N}(y_{1},y_{2})$ can be decomposed onto a basis set of normalized Legendre polynomials $T_{n}(y)=\sqrt{n+\frac{1}{2}}P_{n}(y/Y)$ in the (pseudo)rapidity range of $[-Y,Y]$ as \cite{4,5}:
\begin{equation}
       C_{N}(y_{1},y_{2}) =1+\sum\limits_{n,m=1}^{\infty}\langle a_{nm}\rangle\frac{T_{n}(y_{1})T_{m}(y_{2})+T_{n}(y_{2})T_{m}(y_{1})}{2} .
\label{formula3}
\end{equation}
The first diagonal coefficient $\langle a_{11}\rangle$ represents the contribution from the forward-backward asymmetry of the participating nucleons, $\langle a_{22}\rangle$ reflects the fluctuations in the width of the single particle density distribution, and $\langle a_{nm}\rangle$ coefficients with m$=$n+2 and larger quantify the strength of shorter range correlations \cite{4}.

Two-particle correlation functions for the most central (0-5\%) Au$+$Au events were calculated as the ratio of the same-event and mixed-event pairs,
where the mixed event pairs were formed by sampling from the single particle distributions in (y, $\varphi$, p$_{\rm T}$). Charged pions, kaons (0.2$<$p$_{\rm T}$$<$2~GeV/c) and protons (0.4$<$p$_{\rm T}$$<$2 GeV/c) were studied in this analysis. The reconstructed tracks with $|$y$|$$<$0.5 (0.7 for protons), and $|\Delta y|$$<$1.0 (1.4 for protons) were used. Every track used in the analysis was directly identified as either $\pi$, K, or p by requiring correct values of both the ionization energy loss and time of flight in the STAR TPC and TOF, respectively. Thus, ``charged hadrons" in this analysis, labelled h$^\pm$, are not simply all reconstructed TPC tracks but are doubly directly-identified $\pi$, K, or p particles. Pseudo-correlations caused by the experimental effects of z-vertex smearing were corrected. The systematic uncertainties were estimated by varying the track selection cuts. In addition, track crossing effects \cite{7b} were corrected by p$_{\rm T}$-ordering the particles in each pair and then removing the pairs with $|\Delta y|$$<$0.05 and $-$60$^{\circ}$$<$$|\Delta\varphi|$$<$0$^{\circ}$ from both the same-event and mixed-event pairs.  

In this analysis, the three-dimensional correlation function $R_{2}(y_{1},y_{2},\Delta\varphi)$ was formed to allow the study of both R$_{2}$(y$_{1}$,y$_{2}$), ({\it i.e.} Eq. \ref{formula1}) and R$_{2}$($\Delta$y,$\Delta\varphi$) correlations. The averaged one-dimensional correlation function $\langle$$R_{2}(\Delta y)$$\rangle$ was obtained by projecting the two or three-dimensional correlation function. The same analysis was also done for events from the UrQMD model \cite{6b}, version 3.4.

\section{Results}
\begin{figure}[h]
\centering
 \includegraphics[trim= 0cm 0.25cm 0cm 1.5cm,clip,width=.57\textheight,height=.33\textheight]{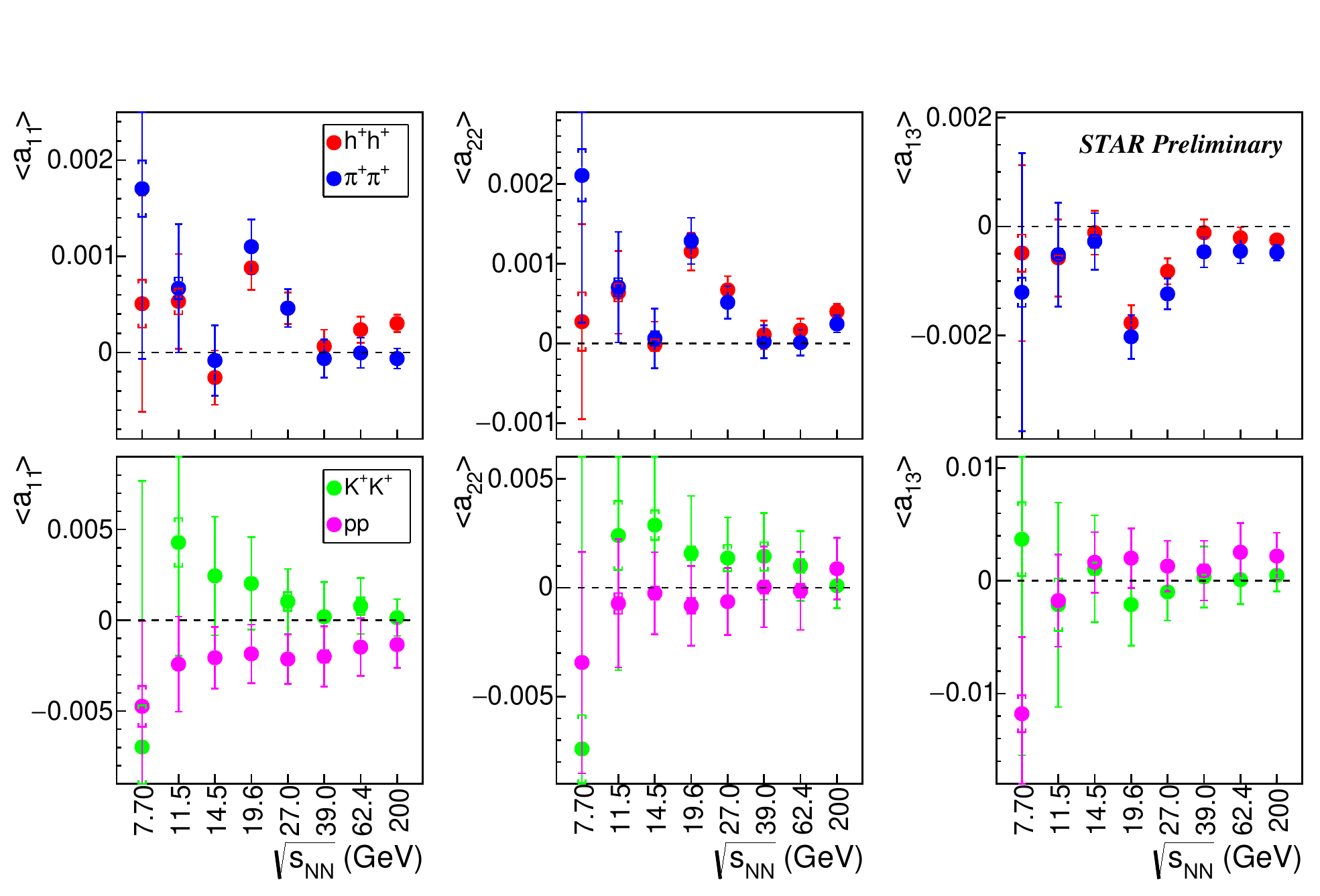}
        \caption{\small Legendre coefficients $\langle a_{11}\rangle$ (left), $\langle a_{22}\rangle$ (middle) and $\langle a_{13}\rangle$ (right) calculated from the two-particle rapidity correlation function for charged hadrons and pions (top), and kaons and protons (bottom), at 8 different beam energies in Au$+$Au collisions. Here, charged hadrons are directly-identified pions, kaons, and protons. The statistical uncertainties are shown as the vertical lines, while the horizontal caps indicate the systematic uncertainties.}
\label{Fig1}
\end{figure}

\begin{figure}[h]
\centering
 \includegraphics[trim= 0.1cm 0.22cm 0.1cm 6.2cm,clip,width=.41\textheight,height=.16\textheight]{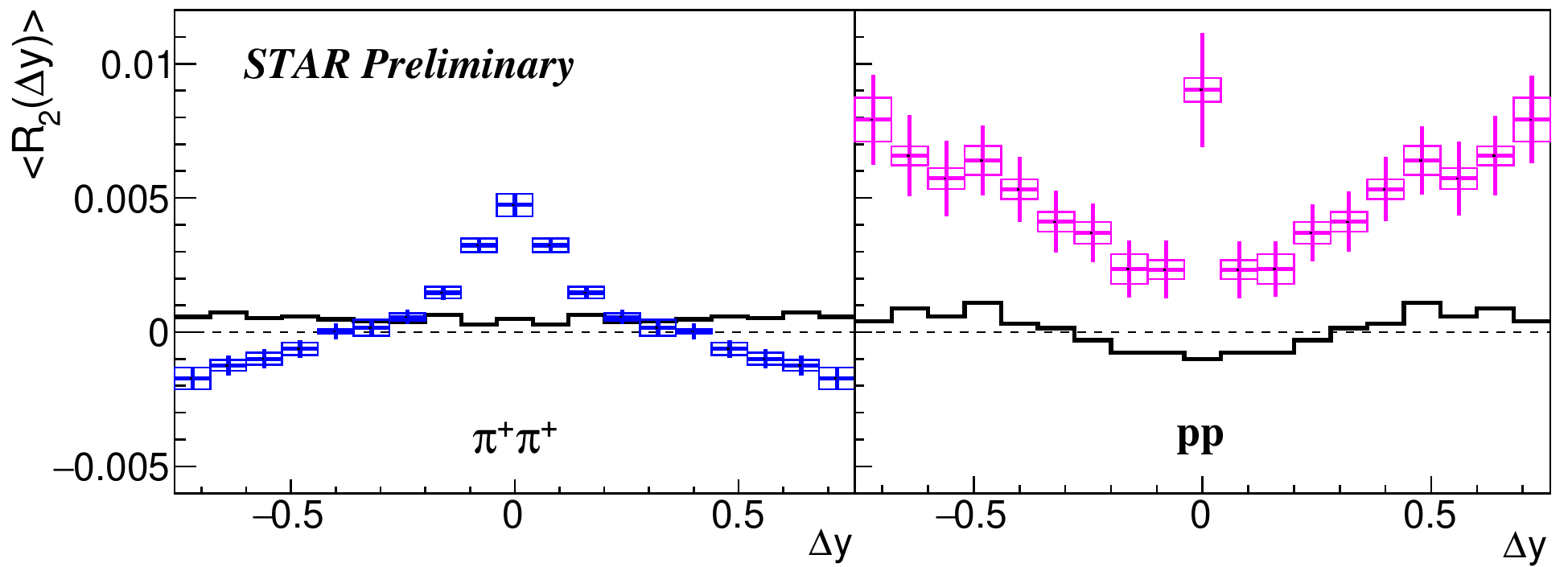}
        \caption{\small The averaged rapidity correlation function for pions (left) and protons (right) over the full $\Delta\varphi$ range at 19.6~$\mathrm{GeV}$. The solid histograms depict the UrQMD results. The middle point at $\Delta$y$=$0 for protons has large fluctuations due to the track crossing effects which exist solely in this bin. }
\label{Fig2}
\end{figure}

Figure \ref{Fig1} shows the most significant diagonal Legendre coefficients $\langle a_{nm}\rangle$ for charged hadrons, pions, kaons and protons for 0-5\% centrality and for each of eight beam energies from 7.7 to 200~GeV.  
The magnitude of the coefficients decreases with increasing beam energy, indicating decreased two-particle correlations per pair as $\sqrt{s_{NN}}$ increases. The first diagonal coefficient $\langle a_{11}\rangle$ is positive for charged hadrons, pions, and kaons. This is expected from wounded nucleon model arguments \cite{3}. However, the $\langle a_{11}\rangle$ coefficient is negative for protons at all eight energies. 
This results from the relative anti-correlation of $\langle$$R_{2}(\Delta y)$$\rangle$ at $\Delta$y$=$0 
that is seen in Figure \ref{Fig2}. 
Such an anti-correlation for protons has also been observed in p$+$p collisions at 7~TeV by ALICE \cite{8} and in $e^{+}e^{-}$ collisions at 29~GeV by the TPC/Two-Gamma collaboration \cite{9}. \par 
A significant increase in the magnitude of the $\langle a_{nm}\rangle$ coefficients for pions at 19.6 and 27.0~$\mathrm{GeV}$ was observed and shown in Figure \ref{Fig1}. A further investigation shows that there is a strong correlation structure in pions around $\Delta$y$\sim$0 that is elongated in the $\Delta\varphi$ direction. This structure only appears for pions and is similar in shape to those observed in cluster emission models applied to p$+$p collisions at RHIC and the LHC \cite{10,11}.\par

\begin{figure}[h]
\centering
 \includegraphics[trim= 2cm 0.45cm 2cm 0.3cm,width=.45\textheight,height=.39\textheight]{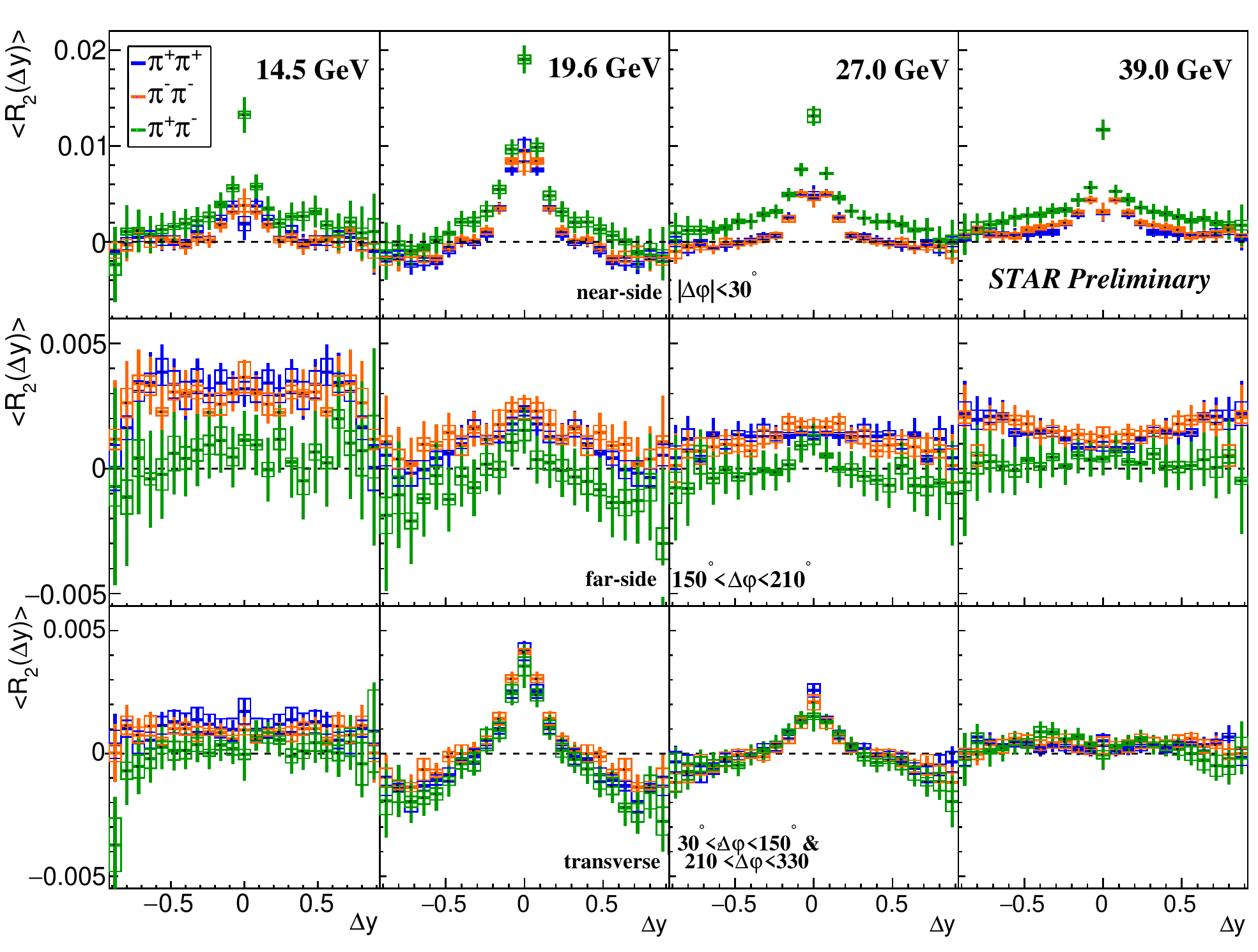}
        \caption{\small The averaged correlation function for pions as a function of $\Delta y$ for 14.5, 19.6, 27, and 39~GeV (from left to right respectively) beam energies in 0-5\% central Au$+$Au collisions. Averages were taken over three different ranges of $\Delta\varphi$: (top) near-side, (middle) far-side and (bottom) transverse region. See the text for the definition of the different $\Delta\varphi$ ranges.}
\label{Fig3}
\end{figure}

In order to further understand the observed structure, the averaged correlation functions for pions versus $\Delta y$ were calculated in three different ranges of the azimuthal angle $\Delta\varphi$: near-side (-30$^{\circ}$$<$$\Delta\varphi$$<$30$^{\circ}$), far-side (150$^{\circ}$$<$$\Delta\varphi$$<$210$^{\circ}$) and transverse $\Delta\varphi$ (30$^{\circ}$$<$$\Delta\varphi$$<$150$^{\circ}$ \& 210$^{\circ}$$<$$\Delta\varphi$$<$330$^{\circ}$) for 19.6 and 27~GeV and their neighboring energies (14.5 and 39~GeV) as shown in Figure \ref{Fig3}. In the near-side projection there is a peak around $\Delta$y$\sim$0 which is stronger in unlike-sign pions than like-sign pions. This is an expected feature of the short-range correlation mechanisms \cite{5} that are dominant in this $\Delta\varphi$ range.
There is no significant structure in the far-side projection of both like-sign and unlike-sign pions at these energies. However, the transverse projection shows a significant peak around $\Delta$y$\sim$0 at 19.6 and 27~GeV. This structure has the same magnitude for both like-sign and unlike-sign pions. The UrQMD model does not reproduce this observation (see Figure~\ref{Fig2}).

\section{Conclusions}
Two-particle rapidity correlations have been studied for like-sign and unlike-sign charged hadrons and directly-identified particles ($\pi, K$, and $p$) in 0-5\% central Au$+$Au collisions at $\sqrt{s_{NN}}$$=$7.7-200~GeV. The shape of the rapidity correlations has been quantified by decomposing the correlation functions onto a basis set of Legendre polynomials. The $\langle a_{11}\rangle$ coefficients have been observed to be positive for pions and kaons, and they are negative for protons at all eight beam energies. The positive or negative values of $\langle a_{11}\rangle$ are indicative of correlations or anti-correlations, respectively. 
A charge-independent structure has been observed in the correlation function for pions. This structure is very localized in beam-energy (19.6 and 27~GeV) and at $\Delta$y$\sim$0 and extends as a ridge in the $\Delta\varphi$ direction. Further studies are needed to understand the underlying physical mechanisms for this structure. 



\bibliographystyle{elsarticle-num}
\bibliography{<your-bib-database>}



\end{document}